\documentclass{80SA}
\usepackage{txfonts,color}
\usepackage{times}

\newcommand{\imag}{\mathrm{Im}\,}

\newcommand{\imu}{\mathrm{i}}

\title{
Electronic Orders Induced by Kondo Effect in Non-Kramers $f$-Electron Systems
}

\author{Yoshio \textsc{Kuramoto}\thanks{E-mail: kuramoto@cmpt.phys.tohoku.ac.jp}, 
Shintaro \textsc{Hoshino} and
Junya \textsc{Otsuki}  
}
\inst{Department of Physics, Tohoku University, Sendai 980-8578, Japan}

\abst{
This paper clarifies the microscopic nature of the staggered scalar order, 
which is specific to  
even number of $f$ electrons per site.
In such systems,  
crystalline electric field (CEF) can make a
singlet ground state.
As exchange interaction with conduction electrons increases, the CEF singlet at each site gives way to Kondo singlets.  The collective Kondo singlets are identified with
itinerant states that form energy bands.  
Near the boundary of itinerant and localized states, a new type of electronic order appears with staggered Kondo and CEF singlets.
We present a phenomenological three-state model that qualitatively reproduces the characteristic phase diagram, 
which have been obtained numerically
with use of the continuous-time quantum Monte Carlo combined with the dynamical mean-field theory. 
The scalar order observed in PrFe$_4$P$_{12}$ is ascribed to this staggered order accompanying charge density wave (CDW) of conduction electrons.
Accurate photoemission and tunneling spectroscopy should be able to probe sharp peaks below and above the Fermi level in the ordered phase.
}

\kword{
Kondo effect,
crystalline electric field singlet,
charge density wave,
dynamical mean-field theory,
quantum Monte Carlo,
PrFe$_4$P$_{12}$
}

\begin{document}
\maketitle

\section{Introduction}

In interacting electron systems with two $f$-electrons per site as in some Pr and U compounds, 
the distinction between 
itinerant and localized characters of electrons
is not as explicit as in Ce or Yb compounds with $f^1$ or $f^{13}$ configuration, respectively. 
The difference comes from the Kramers degeneracy;
 the ground state of each non-Kramers ion such as Pr$^{3+}$ or U$^{4+}$ can be a singlet even without Kondo effect
provided the CEF favors the non-degenerate ground state.   
On the contrary, the Kramers degeneracy in $f^1$ or $f^{13}$ can only be removed by exchange (Kondo) interaction with conduction electrons.
The dichotomy between itinerant and localized characters of $f$-electrons leads to intriguing phenomena in heavy fermion systems.
In particular, mysterious ordering phenomena have been found in 
systems such as PrFe$_4$P$_{12}$ and URu$_2$Si$_2$\cite{kuramoto09, hassinger08}.
In the case of PrFe$_4$P$_{12}$, the order keeps the point-group symmetry around each Pr, and is called the scalar order \cite{kiss06,sakai,kikuchi07}.  

A typical situation in $f^2$ systems 
is the case where the ground state is the CEF singlet.
If some excited CEF levels are involved by interaction with conduction electrons, the ground state becomes non-trivial. 
Let us consider the case where CEF states form
 a quasi-quartet composed of a singlet ground state and a first-excited triplet.
We represent the singlet-triplet levels 
at each site $i$
in terms of two pseudo spins \cite{shiina04,otsuki05}
$\mib{S}_{\gamma i}$ with $\gamma=1, 2$.
Correspondingly, we introduce two bands 
for conduction electrons that 
interact with pseudo spins with the same orbital index $\gamma$
by the exchange interaction $J  > 0$.
If $f$ electrons interact strongly with conduction electrons, the ground state can be a collective Kondo singlet involving CEF triplet at each site.
In this case, $f$ electrons acquire itinerancy because of the Kondo effect.
Competition between these two singlets, in other words between itinerant and localized characters of $f$-electrons, gives rise to rich physics with exotic ordered phases.

The same kind of competition has in fact appeared already in 
the two-impurity Kondo problem \cite{affleck} 
where two $f$ electrons are spatially separated. 
In our case the two $f$ electrons are on the same atomic site.
There are also many theoretical discussions in $f^2$ impurity systems with the CEF singlet \cite{kuramoto92,shimizu95, koga96, hattori05, hoshino09}.
In considering possible ordering in non-Kramers lattice,
we can draw analogy from the two-impurity Kondo problem and the $f^2$ impurity problem.

In this paper, we provide 
intuitive picture how the competition between CEF and Kondo singlets leads to the new electronic order
that has been proposed recently \cite{hoshino10a}.
This novel electronic order has originally been studied by
elaborate numerical work that combines the continuous-time quantum Monte Carlo method (CT-QMC)\cite{rubtsov05, werner06, otsuki07} and
the dynamical mean-field theory (DMFT)\cite{georges96}, which is 
extended for two sublattices corresponding to staggered order \cite{peters07, hoshino10}.
In addition to reviewing some representative physical quantities
that characterize the staggered order,
we introduce in this paper
a toy model with three states per site, and clarify the nature of
the phase diagram, including the tricritical point.
We also suggest a new experiment that can test whether our microscopic picture applies to the actual electronic order in PrFe$_4$P$_{12}$.
According to our scenario, each conduction band
shows a pair of sharp features that develop in the staggered singlet order.

\section{Analogy to two-impurity Kondo problem}

In order to describe the characteristics of our model, we first draw analogy to the two-impurity Kondo problem.
There are two extreme cases concerning the magnitude of the intersite exchange $I$ and the Kondo exchange $J$ both of which are taken to be positive (antiferromagnetic): 
(i) In the case of $I\gg J$, the impurity spin pair form the singlet, and $J$ acts as weak perturbation; 
(ii) In the opposite case $I\ll J$, each impurity spin forms the Kondo singlet by trapping a conduction electron.  Then $I$ acts as weak perturbation.
In the competing case $I\sim J$, the energy difference between the two kinds of singlets becomes smaller than either of $I$ or $J$.   The difference becomes simply $|J-I|$ if the width of conduction band is negligible.
It is known from detailed analysis \cite{affleck} that the two different singlets have the anti-crossing around $J\sim I$ in general, but they become degenerate if the particle-hole symmetry is present in the conduction band.  In the former case, as the Kondo exchange increases, the localized character of $f$ electrons, represented by the pair singlet, changes continuously to itinerant character, represented by two Kondo singlets. 

We regard the $f^2$ configuration in Pr$^{3+}$ or U$^{4+}$ ion as the short-distance limit of two Kondo centers.   
Instead of two real spins that tend to form the triplet by the strong intra-atomic exchange, we consider two pseudo spins that have
antiferromagnetic exchange interaction $I\ (>0)$.  
Then the triplet level is located higher by $I=\Delta$ relative to the singlet.
The relevant Pr and U compounds are modeled as
the periodic lattice of these $f^2$ pseudo spins.

\section{Generalized Kondo lattice with two conduction bands}

Although the actual Pr and U compounds have complicated band structures and interactions, we consider the simplest model that contains the essential feature of the competition between itinerant and localized characters.  We assume the presence of two conduction bands indexed by $\gamma \ (=1, 2)$ each of which interacts with a localized pseudo spin 
$\mib{S}_{\gamma i}$ at site $i$.
The model is given by
\begin{align}
{\cal H} & = 
   \sum _{\mib{k}\gamma \sigma} ( \varepsilon _{\mib{k}\gamma} - \mu ) c_{\mib{k}\gamma\sigma} ^\dagger c_{\mib{k}\gamma\sigma}
 + 
J \sum _{i \gamma}  \mib{S}_{\gamma i} \cdot \mib{s}_{{\rm c} \gamma i}    \nonumber \\
 & + 
\Delta  \sum_{i}\mib{S}_{1i} \cdot \mib{S}_{2i},
\label{eq_hamilt}
\end{align}
where 
$\mib{s}_{{\rm c}\gamma i}$ denotes the spin of conduction electrons at site $i$.
The CEF splitting $\Delta$ is simulated by
the coupling between pseudo spins
$\mib{S}_{1i}$ and $\mib{S}_{2i}$
 as shown in the third term. \cite{shiina04,otsuki05}
This model at half filling has been investigated in one dimension by the density-matrix renormalization group\cite{watanabe99}.
In this paper we concentrate on the case where each site has one conduction electron on the average.  This means that each conduction band is quarter-filled if the bands $\gamma =1,2$ are equivalent. 

We use the dynamical mean-field theory (DMFT) for setting up an effective impurity model, and employ the continuous-time quantum Monte Carlo (CT-QMC) to solve the impurity model.
In numerical calculation, 
the density of states of the conduction band is taken to be
\begin{align}
\rho_c (\varepsilon) = \frac{1}{D} \sqrt{\frac{2}{\pi}} \exp \left[ 
-2 \left( \frac{\varepsilon}{D} \right) ^2 \right],
\end{align}
where the origin of energy is at the center of the band.
The density of states corresponds to the tight-binding model in the hypercubic lattice in infinite dimensions.
Unless otherwise stated, the parameters to be used in the following numerical results are taken to be: 
$D=1, J=0.8, \Delta=0.2$.
The Kondo temperature in the single impurity Kondo model becomes
$T_{\rm K} \sim 0.1$ according to the formula
$
T_{\rm K} = D\exp\left\{ 
 -1/\left[ 
 J\rho_c (\mu)) \right] \right\} 
$

\begin{figure}
\begin{center}
\includegraphics[width=85mm]{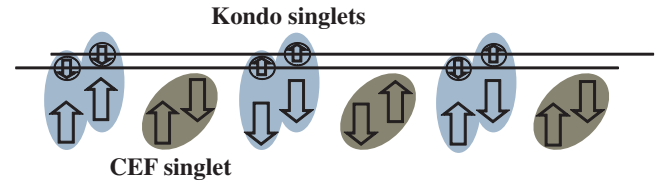}
\end{center}
\caption{(Color online) 
Illustration of the ordering pattern in the strong coupling limit.
The thick arrows represent localized spins, while objects on the two lines represent conduction electrons with spin.
}
\label{pattern}
\end{figure}

Since the method of calculation has been explained in detail in our previous papers \cite{otsuki09, hoshino10a}, we do not repeat the explanation here.
Instead, we describe qualitatively the ground state in the strong-coupling limit.
Figure \ref{pattern} illustrates the spin and charge densities in the ordered phase.
As in the two-impurity Kondo problem, the competition between $J$ and $\Delta$ in eq.(\ref{eq_hamilt}) gives rise to a small energy scale $|J-\Delta|$ in the strong-coupling limit.
If the attractive intersite interaction between these two kinds of singlets exceeds the difference $|J-\Delta|$, the staggered order becomes
energetically favorable.
By precise numerical work that fully includes the kinetic energy of conduction electrons,
we have confirmed that this staggered order is indeed realized with certain set of parameters.
In the following we describe the salient features of this new kind of order.

\section{Phase diagram}

In our calculation the whole lattice is divided into sublattices A and B.
The most fundamental quantity is the site-diagonal component 
$G_{\gamma\sigma} ^{\lambda } (\imu \varepsilon _n)$ of Matsubara Green function with additional sublattice index $\lambda=$ A, B.
We first derive the Matsubara Green function using CT-QMC plus DMFT, 
and then other quantities such as the density of states by analytic continuation to real frequencies.  
If there is no electronic order,  local Green functions do not depend on sublattices.
The phase transition is the point where the Green function for the A sublattice begins to be different from that for B.  
Another important quantity is the two-particle Green function of
conduction electrons, and that of $f$ electrons representing localized spins.
From the two-particle Green function with imaginary frequencies,  we obtain the dynamical susceptibilities by analytic continuation to real frequencies.  
On the other hand, 
spin correlation functions are obtained by direct evaluation of the statistical average.

\begin{figure}
\begin{center}
\includegraphics[width=85mm]{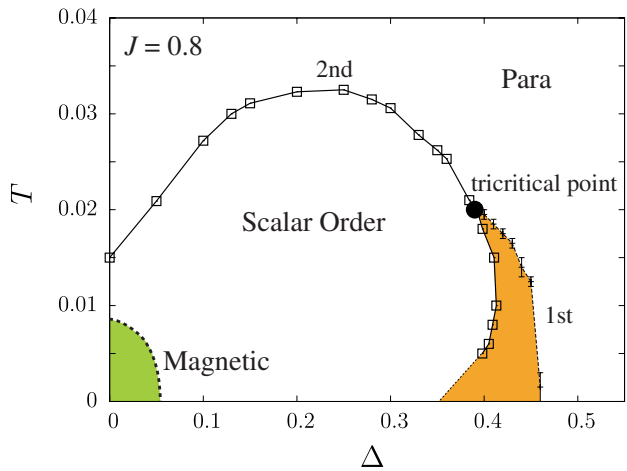}
\caption{
Phase diagram 
with one conduction electron per site. Here and in all numerical results,
the exchange interaction is set to $J=0.8$.
The tricritical point separates the second- and first-order transitions.  Estimated upper and lower metastability limits in the first-order transition are indicated.  See text for the phase labeled "Magnetic".
}
\label{fig_phase_d}
\end{center}
\end{figure}

A second-order transition is signaled by 
divergence of a susceptibility that corresponds to the response of the order parameter.  In our case,
an element of 
the staggered charge susceptibility $\chi_{\rm c}$, 
which has symmetric orbital combination
diverges at the highest $T$.
We represent this element as $\chi _{+ \rm c}(\mib{Q})$.
Figure \ref{fig_phase_d} shows the phase diagram obtained from the divergence of
$\chi _{+ \rm c}(\mib{Q})$.
There appears a region where the first-order transition occurs.
This is detected from the hysteresis in physical quantities as we vary $T$ or $\Delta$.  
For example, 
as $\Delta$ is decreased from $\Delta=0.5$ at $T=0.015$,
$1/\chi _{+ \rm c}(\mib{Q})$ tends to zero around $\Delta=0.41$.
On the other hand, starting from the ordered phase,
$1/\chi _{+ \rm c}(\mib{Q})$ 
is finite at $T=0.015$ and $\Delta=0.41$, and decreases to zero around
$\Delta=0.44$.
Hence, the intervening region $0.41<\Delta<0.44$ corresponds to metastable state at $T=0.015$. 
The metastable region shrinks with increasing $T$, and terminates at the tricritical point at $T\sim 0.2$.
Because of the numerical difficulty, the phase boundary around $\Delta \sim 0.46$ has been obtained only crudely. 
  
We find that the homogeneous susceptibility for the quantity
\begin{align}
\sum_{\sigma=\pm 1}\sum_{\gamma }
\sigma 
\tau_\gamma 
n_{\rm c \gamma\sigma},
\end{align}
with $\tau_1=1$ and $\tau_2=-1$
diverges at $T < 0.01$ inside the scalar-ordered phase.
The order parameter breaks the time reversal even without finite magnetization.  
Hence this order parameter is interpreted as a kind of magnetic octupole. 
Detailed study of this order will be reported elsewhere.

Let us introduce a phenomenological model that is convenient to interpret the phase diagram.
Taking the strong-coupling limit, we consider the following three states for each site $i$:
(i) the Kondo state with two conduction electrons with energy $E_2$;
(ii) the CEF singlet state with no conduction electrons with energy $E_0$;
(iii) the CEF singlet state with one conduction electrons with energy $E_1$,
which is taken to be 0.  
The states (i) and (ii) are nondegenerate, while
the state (iii) is four-fold degenerate 
since the conduction electron has spin and orbital degrees of freedom, 

The variable $\sigma_i$ is assigned for each state with a value:  
(i) $\sigma_i = 1$;
(ii) $\sigma_i = -1$;
(iii) $\sigma_i = 0$.
Furthermore, the constraint
\begin{align}
\sum_{i=1}^N \langle \sigma_i \rangle = 0,
\label{constraint}
\end{align}
is imposed to describe the situation of one electron per site on the average. 
This constraint is accomplished by adding the Lagrange multiplier term
$h \sum_i \sigma_i$ to the Hamiltonian.
In the mean-field theory, we obtain the partition function $Z_\lambda$ and the 
average $\langle \sigma\rangle_\lambda$
for each sublattice $\lambda=$A, B.
Assuming 
$\langle \sigma\rangle_{\rm A} \equiv \sigma_{\rm A} 
\ge 0$ and
$\langle \sigma\rangle_{\rm B} \equiv \sigma_{\rm B} \le 0$,
we obtain for $\lambda =$A,
\begin{align}
Z_{\rm A} &= 
\exp\left[ -\beta (E_2+ h +K\sigma_{\rm B}) \right]+ 4\nonumber\\
&+
\exp\left[ -\beta (E_0-h- K \sigma_{\rm B}) \right],
\label{ZA}
\end{align}
where $K\ (>0)$ is the interaction between the nearest neighbor sites,
and is given in terms of the hopping $t$ as
$K\sim O(z t^2/\delta E)$ where $z$
is the number of nearest neighbors.
The excitation energy for the state (iii) is written as $\delta E$, which is 
of the order of $J$ or $\Delta$.
Note that $K$ describes the effective attraction between the states (i) and (ii).
We obtain similar result for $Z_{\rm B}$.

We parameterize as $E_2 = E +d$ and $E_0 = E -d$ where $2d$ is estimated as
$\Delta -J$ in the strong-coupling limit.
The constraint of eq.(\ref{constraint}) requires $h=-d$, and eq.(\ref{ZA}) is simplified to
\begin{align}
Z_{\rm A} &= 
2\exp\left( -\beta E \right) \cosh (\beta K\sigma) +4,
\end{align}
where we have used 
the constraint 
$\sigma_{\rm A} =-\sigma_{\rm B}\equiv \sigma$.
The partition functions of both sublattices become equal to each other: 
$Z_{\rm A} = Z_{\rm B} \equiv Z$.
Let us consider some limiting cases,
which clarify the meaning of $E$.  
In the limit $E \rightarrow -\infty$, 
our model is reduced to the 
antiferromagnetic Ising model.   
This is understood as exclusion of the state (iii) with $n_{\rm c}=1$.
In the opposite limit of $E/K\gg 1$, all sites become (iii), and no staggered order occurs.  Hence, the value of $E/K$ controls 
the relative weight of the state (iii), and subsequently 
the phase transition. 

The free energy $F$ per site is given by
\begin{align}
F = -T\ln Z +\frac 12 K \sigma^2.
\end{align}
The equation of state is obtained from the stationary condition $\partial F/\partial \sigma =0
$ as
\begin{align}
\sigma  = \frac 2{Z}\exp(-\beta E)\sinh 
(\beta K\sigma) \equiv g(\sigma).
\label{sigmaA}
\end{align}
Nontrivial solution $\sigma\neq 0$ is possible
only if $E<K$.

In the second-order transition, 
the nontrivial solution becomes infinitesimal at $T_c$, which tends to $T_c = K$
in the limit $E\rightarrow -\infty$.
As $E$ becomes larger,  the state (iii) acquires a finite weight.
The transition temperature $T_c$ is then given by
\begin{align}
T_c = \frac {K}{1+2\exp(\beta_c E)},
\end{align}
which is a transcendental equation for $T_c = 1/\beta_c$.  The resultant $T_c$ is smaller than $K$.

The second-order transition beginning from 
$E$ from $-\infty$ encounters a 
tricritical point at $E=E_3$ with transition temperature $T=T_3$.
For larger $E$, the transition becomes of first order.
In order to derive the parameters corresponding to this tricritical point, we expand the RHS of eq.(\ref{sigmaA}) up to the third order in $\sigma$.  
The coefficient of the third-order term becomes zero at the tricritical point.  
We derive the parameters as
\begin{align}
E_3 =0, \quad 
T_3 = K/ 3.
\end{align}
The first-order transition 
eventually disappears for larger $E$ at $E=E_c$.  We obtain
$ E_c = K$ by considering the low-temperature limit of eq.(\ref{sigmaA}).  
For $E>E_c$
the model has no phase transition.  

\begin{figure}
\begin{center}
\includegraphics[width=85mm]{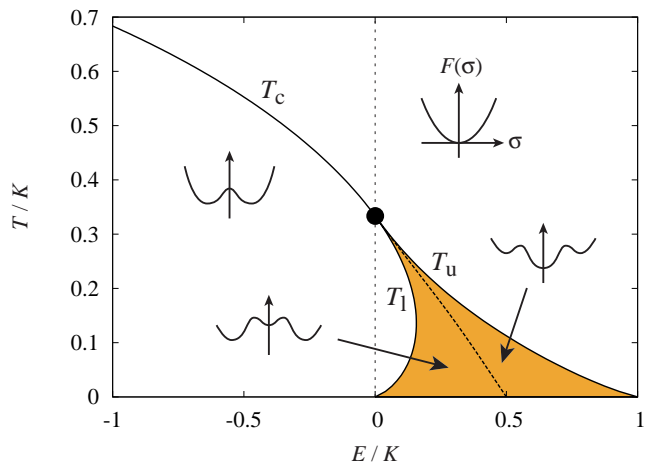}
\end{center}
\caption{(Color online) 
Phase diagram of the three-state model, together with illustration of the free energy as a function of the order parameter $\sigma$.   
As $E$ increases, the second-order transition changes to first-order one beyond the tricritical point at $(E,T)=(0,K/3)$.  The limits 
of metastability 
appear on both sides ($T_u$ and $T_l$) of the thermodynamic transition line.}
\label{3state}
\end{figure}

Figure \ref{3state} shows the phase diagram of the three-state model obtained numerically.
The temperature 
$T_u$ giving the upper limit of the first-order transition
is determined by the condition 
\begin{align}
g' (\sigma) =1,
\end{align}
for $\sigma\neq 0$.  
The lower limit $T_l$ is given by the same equation but with $\sigma=0$.
As shown in Fig.\ref{3state}, the lower metastability limit appears mostly as a reentrant transition.
The thermodynamic transition line corresponds to the condition
\begin{align}
F(\sigma) = F(0),
\label{equili}
\end{align}
where $\sigma\ (\neq 0) $ is the solution of eq.(\ref{sigmaA}).
Since we have $\sigma=1$ with $T\rightarrow 0$, 
eq.(\ref{equili}) is satisfied only if 
\begin{align}
-(K-E)+K/2 =0,
\end{align}
which gives $E/K=0.5$ as shown in Fig.\ref{3state}.

These results are useful to understand the numerical results in Fig.\ref{fig_phase_d}, which are not in the strong-coupling limit.
We interpret $T_{\rm K}$ as playing the role of $J$.
Since increase of $\Delta >0$ causes decrease of $T_{\rm K}$,
the transition temperature increases up to the point where $T_{\rm K}$ becomes comparable to $\Delta$.
In the phenomenological three-state model,  increase of 
$2d=\Delta-T_{\rm K}$, with $E_0  =E-d$ kept fixed, means increase of $E$.
Namely, the increase of $\Delta$ in Fig.\ref{fig_phase_d} corresponds to increase of $E$.
Thus the change of second-order transition to first-order one, and eventual disappearance with increasing $\Delta$ can be interpreted by the three-state model.
It is remarkable that the salient feature of Fig.\ref{fig_phase_d} is reproduced qualitatively by this simple model.

\section{Static quantities in ordered phase}

\subsection{Staggered charge density}

\begin{figure}
\begin{center}
\includegraphics[width=85mm]{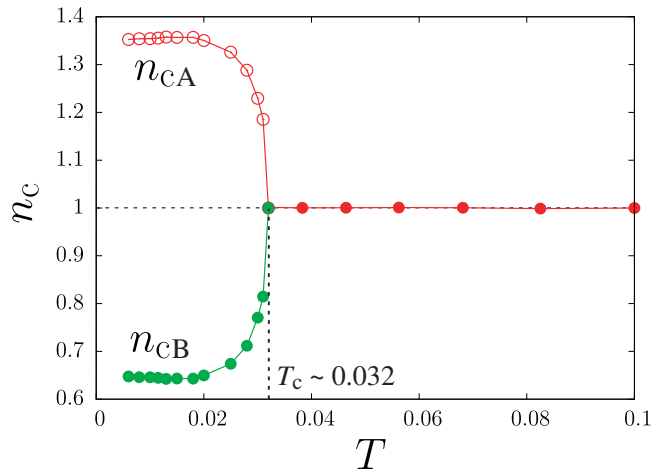}
\end{center}
\caption{(Color online) 
Average occupation number of conduction electrons at each sublattice A or B.
}
\label{CD}
\end{figure}

Figure \ref{CD} shows the number of conduction electrons
per site. 
This quantity has 
been obtained for each sublattice $\lambda = $ A or B 
from
\begin{align}
n_{\rm c\lambda} = T\sum_{\gamma\sigma}\sum_n 
G_{\gamma\sigma} ^{\lambda } (\imu \varepsilon _n).
\end{align}
We have fixed the average number $n_{\rm c}=1$ distributed over the two equivalent conduction bands, as seen 
in Fig.\ref{CD} above the transition temperature  $T_c \sim 0.032$.
As temperature $T$ becomes lower than $T_c$, the disproportion occurs between the sublattices A and B.  
Their sum $n_{c\rm A}+n_{c\rm B}$ remains constant $(=2)$ as in $T>T_c$ by construction.
Hence the electronic order we have obtained corresponds to a 
CDW state of conduction electrons.
The CDW is related to the difference of local spin correlations as shown next.

\subsection{Local spin-correlation functions}

\begin{figure}
\begin{center}
\includegraphics[width=85mm]{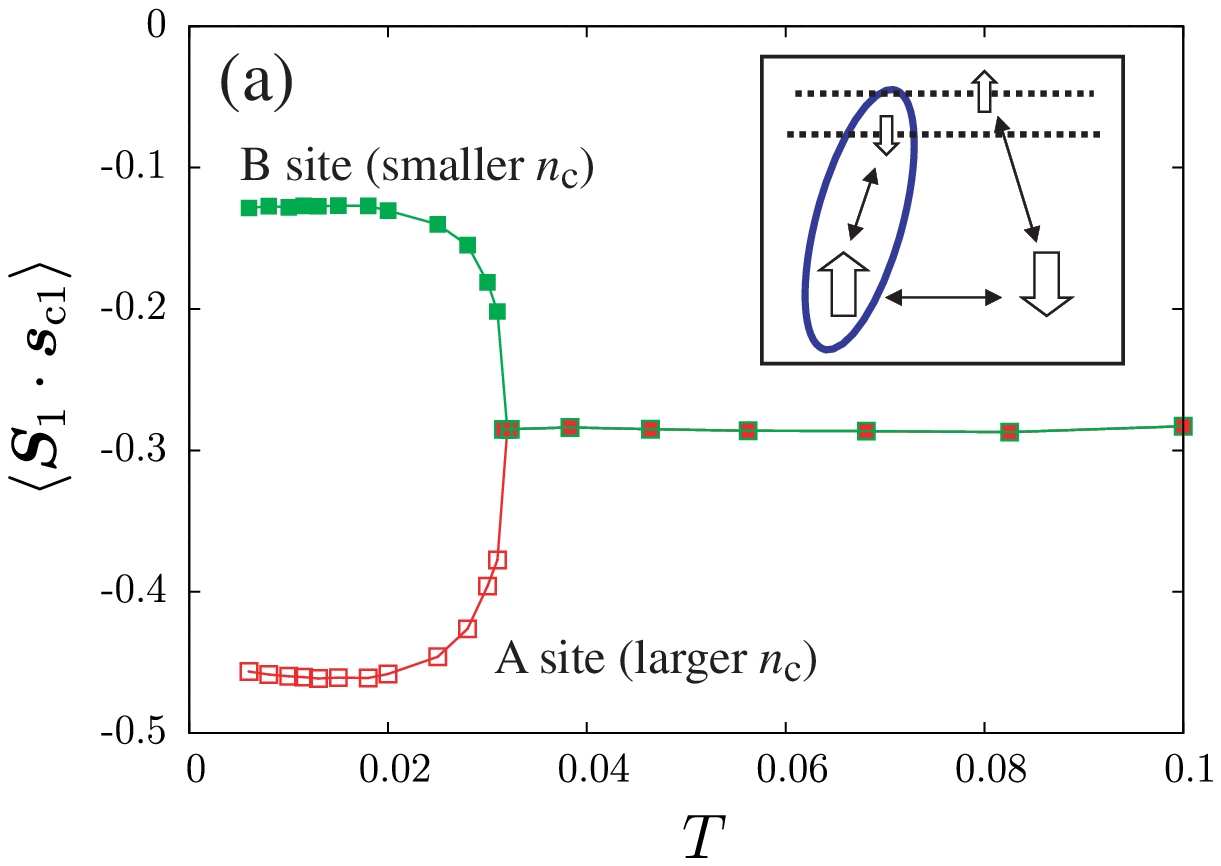}
\includegraphics[width=85mm]{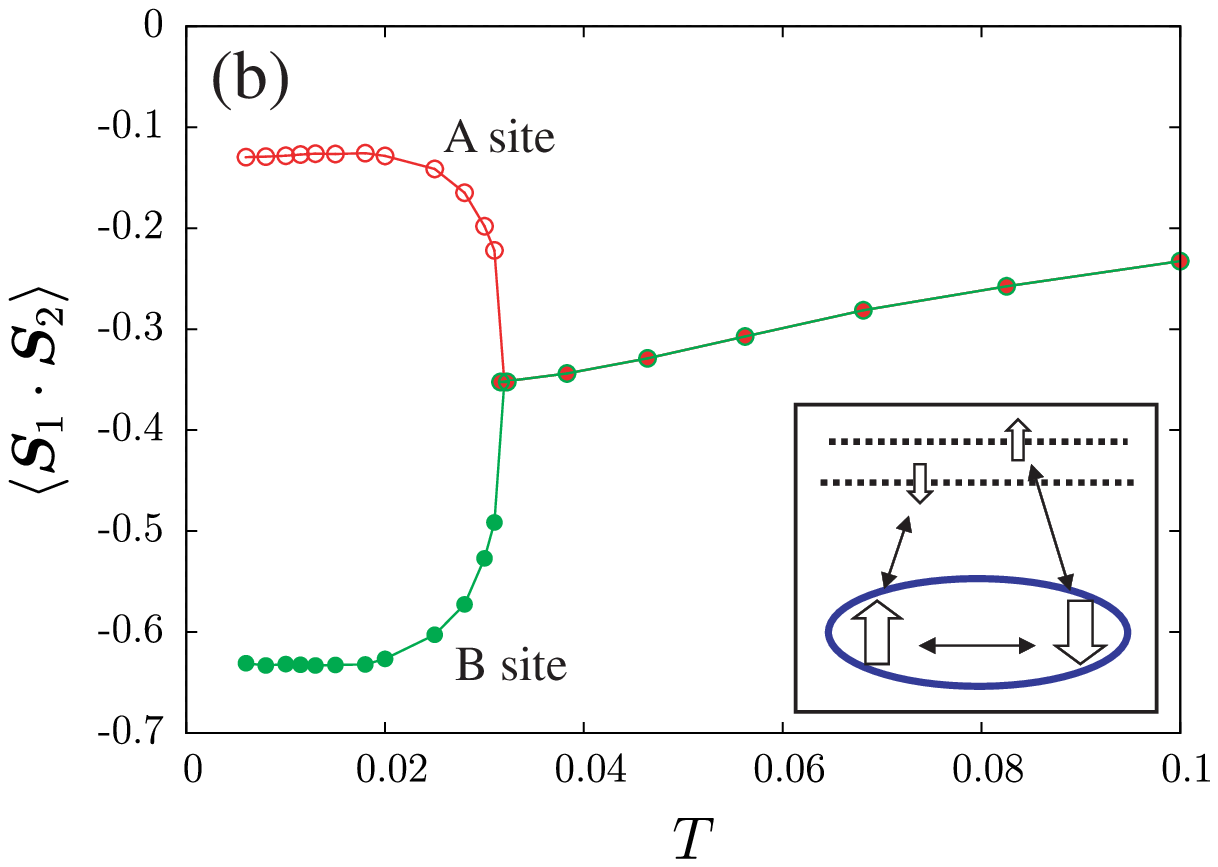}
\end{center}
\caption{(Color online) 
Local correlation functions 
(a) between conduction and localized spins representing the Kondo effect, and
(b) between localized spins representing the stability of the CEF singlet relative to the triplet. 
}
\label{spin-correlation}
\end{figure}

Let us examine equal-time spin correlations that
clarify properties of each sublattice
with the finite CEF splitting.
We first consider 
$\langle \mib{S}_1 \cdot \mib{s}_{\rm c1} \rangle$, which is equal to $\langle \mib{S}_2 \cdot \mib{s}_{\rm c2} \rangle$ under the present condition.
Figure \ref{spin-correlation}(a) shows the temperature dependence of
$\langle \mib{S}_1 \cdot \mib{s}_{\rm c1} \rangle$, which
is enhanced 
on
A-sublattice and suppressed 
on
B-sublattice.
Bearing $n_{\rm cA} > n_{\rm cB}$ in mind, we conclude that the localized spin on A-sublattice forms the Kondo singlet.
On the other hand, the correlation $\langle \mib{S}_1 \cdot \mib{S}_2 \rangle$ between localized spins is enhanced at B site as shown in Fig. \ref{spin-correlation}(b).
Hence, the B-sublattice corresponds to the CEF singlet.

The magnitude of the effective CEF splitting for the CEF-singlet site can be estimated as follows:
Taking the effective Hamiltonian for the localized states as ${\cal H}_f^{\rm eff} = {\tilde \Delta} \mib{S}_1 \cdot \mib{S}_2$,
we obtain 
the susceptibility as $\chi_{\rm M}^{12} = - 1 / (2{\tilde \Delta})$ for the ground state.
Here ${\tilde \Delta}$ is the effective CEF splitting, 
which can be estimated from the local susceptibility in B-sublattice. 
The numerical result is ${\tilde \Delta} = 0.098$, which is very close to the original CEF splitting $\Delta = 0.1$.
Hence,  spatially extended Kondo singlets do not significantly affect the magnitude of the CEF splitting.
Besides, in Figure \ref{spin-correlation}(b), $\langle \mib{S}_1 \cdot \mib{S}_2 \rangle$ in the low-temperature limit is not far from $-0.75$ expected for the isolated singlet.
Hence the CEF singlet interacts only weakly with conduction electrons.
Note, however, that the correspondence between $\Delta$ and $\tilde \Delta$ does not hold for $\Delta \lesssim 0.05$.
This is because the CEF-singlet site tends to be magnetically polarized
near $\Delta = 0$.

In this way, this electronic order 
turns out to be 
a staggered order with the Kondo and CEF singlets, as illustrated in Fig.\ref{pattern}. 
Except for the strong coupling limit, the number of conduction electrons at the CEF singlet site is not zero because the Kondo singlets are spatially extended.  This fact becomes important in understanding the single-particle density of states to be presented in the next section.

\section{Dynamical quantities}

\subsection{Dynamical susceptibility}

In our theoretical framework, it is in principle possible but numerically tedious to 
derive the momentum-dependent dynamical susceptibility.
Hence, prior to full derivation of the dynamical susceptibility in the future,
this paper presents some results only for the local dynamical susceptibility
that corresponds to the momentum average.
In skutterudites such as PrFe$_4$P$_{12}$,
the triplet wave functions $|\Gamma_t\rangle$
are mixture of the cubic $\Gamma_4$ and $\Gamma_5$ states because of the lower local symmetry $T_h$.
Namely the wave function is given by
\begin{align}
|\Gamma_t\rangle = \sqrt{w}|\Gamma_4\rangle +\sqrt{1-w}|\Gamma_5\rangle.
\end{align}
Then matrix elements of the magnetic dipole between singlet and triplet levels 
are fixed for given $w$.  Note that the limit $w=0$ makes the dipole moment vanish.

\begin{figure}
\begin{center}
\includegraphics[width=85mm]{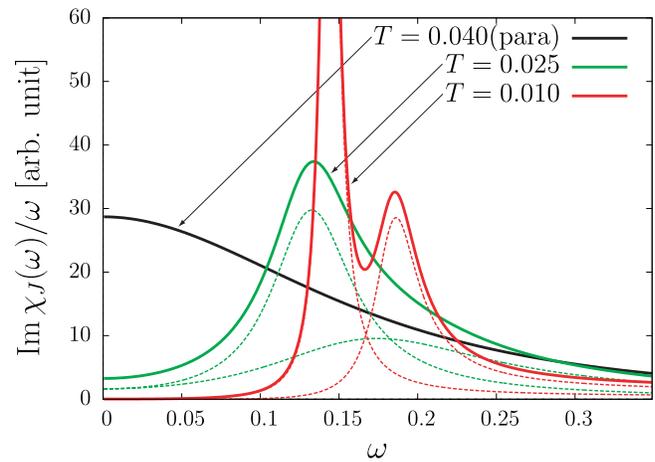}
\end{center}
\caption{(Color online) 
Local dynamical magnetic susceptibility at different temperatures.  Below the transition temperature $T_c\sim 0.032$,  A and B sublattices have different spectra, as shown by thin lines.
The Kondo-singlet site has the broader and higher-energy peak, while the CEF singlet site has the shaper and lower-energy peak as most clearly seen at $T=0.010$.  
}
\label{local-chi}
\end{figure}

Figure \ref{local-chi} shows the results related to the dynamic magnetic susceptibility $\chi_J(\omega)$
above and below the transition temperature.
Here we have used the parameter $w=0.64$ for the numerical calculation.
At $T=0.04$, which is
above the transition temperature $T_c\sim 0.032$, the Lorentzian spectrum emerges characterized by the quasi-elastic components.
As $T$ decreases below $T_c$, the first inelastic feature appears with the peak at
$\omega_1\sim 0.14$.  With further decrease of $T$, the second inelastic peak also develops at $\omega_2\sim 0.18$.  
By looking at sublattice components,  we have identified 
the first inelastic peak at $\omega_1$ comes from the CEF singlet sites, while
the second peak at $\omega_2$ from the Kondo singlet sites.

\subsection{Single-particle spectrum}

\begin{figure}
\begin{center}
\includegraphics[width=85mm]{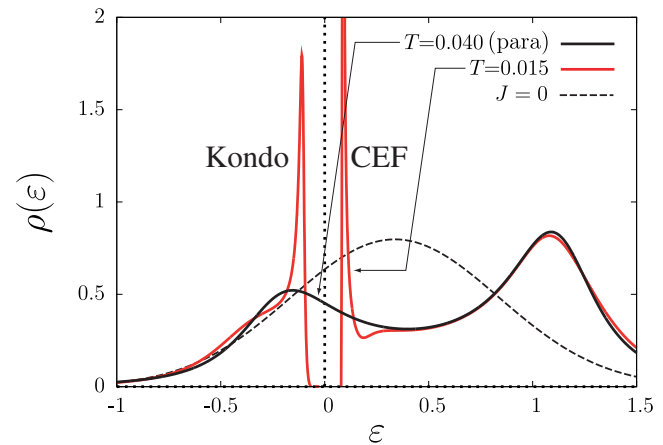}
\end{center}
\caption{(Color online) 
Density of states for conduction electrons at different temperatures.
The Gaussian density of states of non-interacting case ($J=0$) is also shown.  The sharp peak 
below the Fermi level at $T=0.015<T_c$ comes from the Kondo-singlet site, while the other peak above the Fermi level from the CEF singlet site.
}
\label{DOS}
\end{figure}

The momentum average of single-particle dynamics can be seen in the density of states.  
Figure \ref{DOS} shows the renormalized density of states 
given by 
\begin{align}
\rho (\varepsilon) = 
- \frac{1}{2\pi} \sum_{\lambda={\rm A,B}} \sum_{\gamma\sigma}
\imag G^{\lambda}_{\gamma\sigma} (\varepsilon + \imu \delta),
\label{rDOS}
\end{align}
for $J=0.8$ and $\Delta = 0.2$.
Here the local Green function $G^{\lambda}_{\gamma\sigma} (\varepsilon + \imu \delta)$ 
of conduction electrons 
is derived by analytic continuation of Matsubara frequencies.
Under the present condition, the Green functions 
does not depend
on the labels $\gamma$ and $\sigma$. 

In the disordered state ($T=0.04$),
the density of states is finite at the Fermi level, indicating
the metallic behavior.
The large dip structure around $\omega \sim 0.3$ is interpreted as a kind of hybridization
 between conduction electrons and the localized states by the Kondo effect.
Although there is no real hybridization because the pseudo spins do not have charge degrees of freedom, strong renormalization by the Kondo effect gives rise to an electronic state that allows this interpretation
\cite{otsuki09a}. 
At $T=0.015$ which is a temperature inside the ordered phase, the double peaked structure arises as in Kondo insulators.
In the present case, however, each peak is associated with different sublattices.
Namely, 
the sharp peak below the gap comes from the sublattice for the Kondo singlet, while the peak above the gap is due to the CEF singlet site. 
Hence, this 
double peaked structure 
clearly shows the difference of 
the occupation number
between the Kondo and CEF singlet sites.
We emphasize, however, that the Kondo effect is responsible for both peaks, as in the case of Kondo insulators.

The origin of the insulating behavior is explained as follows.
Each conduction band has one conduction electron per doubled unit cell in the ordered state.
Provided that the localized spin at the Kondo-singlet site participates to the conduction band, each band is filled by two electrons.
Then the system can be an insulator.
In this viewpoint, the staggered Kondo-CEF singlet order may be regarded as 
alternating itinerant and localized sites of $f$ electrons.

\section{Discussion and conclusion}

\begin{figure}
\begin{center}
\includegraphics[width=85mm]{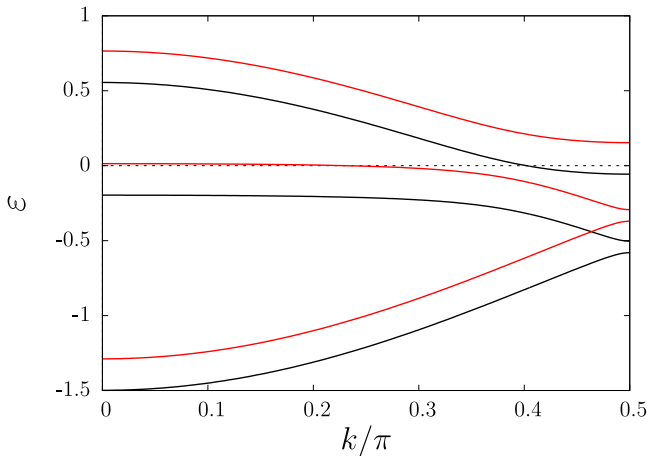}
\end{center}
\caption{(Color online) 
Renormalized energy bands schematically shown in one dimension.  A pair of original conduction bands hybridize with Kondo states, and end up with six energy bands in the reduced Brillouin zone in the ordered phase.
We have used 
$\varepsilon = -\cos k+const.$ for the spectrum of bare conduction bands.  The flat parts in the renormalized bands appear as peaks in $\rho (\epsilon)$.
With one conduction {\it hole} per site, the Kondo-singlet site has a peak above the Fermi level, while the CEF singlet site has another peak
below the Fermi level. This corresponds to the reversal of energy from the result shown in Fig.\ref{DOS}.
Furthermore, the two conduction bands here are not equivalent, as indicated by different colors.
}
\label{bands}
\end{figure}

Let us finally discuss the relevance of the present model to real systems such as PrFe$_4$P$_{12}$.  
We have to consider two important effects in PrFe$_4$P$_{12}$:
(i) the two conduction bands are not equivalent;
(ii) there is one conduction {\it hole}, rather than one conduction {\it electron}, 
 per Pr ion.
As a result of (ii), the CDW gives more holes to 
the Kondo singlet site 
than the CEF site.  
Then the Kondo singlet site has the peak {\it above} the Fermi level
for the local density of states of conduction electrons, 
and the CEF singlet site has the peak {\it below} the Fermi level.  
The ensuing renormalized energy bands are illustrated schematically in Fig.\ref{bands}, where the effect (i) is also taken into account.
The number of itinerant electrons is 8 $(=6+2)$ per unit supercell in the ordered phase, out of which two electrons originate from the Kondo-singlet site with $f^2$ configuration.  Hence the ground state is a semimetal or semiconductor, depending on the degree of asymmetry in conduction bands.

The CDW of conduction electrons should lead to distortion of crystals keeping the local point-group symmetry.  
The distortion has indeed been observed by diffraction experiments by X rays and neutrons\cite{iwasa02,iwasa03,hao05}.
In inelastic neutron scattering on PrFe$_4$P$_{12}$, clear inelastic features appear only below the transition temperature.\cite{iwasa03,park08}
If a small amount of Pr is replaced by La, the scalar order disappears, and no inelastic peaks are visible even at low enough temperatures.\cite{iwasa08}
The results shown in Fig.\ref{local-chi} reproduce the characteristic features observed in PrFe$_4$P$_{12}$, giving further support to relevance of our model.

The clearest confirmation of the present 
scenario will be provided by accurate spectroscopy such as the 
photoemission and tunneling spectroscopy 
in PrFe$_4$P$_{12}$.
Namely, only below the ordering temperature, 
sharp peaks of electronic density of states
develop both below and above the Fermi level.
Photoemission probes
the sharp feature below the Fermi level.
Above the Fermi level, on the other hand, the hybridized heavy-electron band 
may be probed by tunneling spectroscopy, which has been utilized effectively
in URu$_2$Si$_2$.\cite{schmidt} 

In summary, 
we have described the new staggered singlet order for non-Kramers $f$-electron systems, and discussed its salient features in static and dynamic quantities.  
For the phase diagram, 
we have proposed
a simple model and reproduced the first- and second-order transitions depending on the relative stability of CEF and Kondo singlets.  
The tricritical point measures the strength of the intersite interaction 
that is responsible for
the staggered order.
The scalar order observed in PrFe$_4$P$_{12}$ is ascribed to
this singlet order.

This work has been supported by 
a Grant-in-Aid for Scientific Research 
No.20340084, and 
another (No. 20102008, 
Innovative Areas "Heavy Electrons" ) 
of The MEXT, Japan.


\begin{thebibliography}{99}
\bibitem{kuramoto09} For a review, see: Y. Kuramoto, H. Kusunose, and A. Kiss: J. Phys. Soc. Jpn. {\bf 78} (2009) 072001.
\bibitem{hassinger08} 
For comparison between PrFe$_4$P$_{12}$ and URu$_2$Si$_2$, see:
E. Hassinger {\it et al}: 
J. Phys. Soc. Jpn. 77 (2008)  Suppl. A, p. 172.
\bibitem{kiss06} A. Kiss and Y. Kuramoto: J. Phys. Soc. Jpn. {\bf 75} (2006) 103704.
\bibitem{sakai} O. Sakai {\it et al}: J. Phys. Soc. Jpn. {\bf 76} (2007) 024710.
\bibitem{kikuchi07} J. Kikuchi, M. Takigawa, H. Sugawara, and H. Sato: J. Phys. Soc. Jpn. {\bf 76} (2007) 043705.
\bibitem{shiina04} R. Shiina: J. Phys. Soc. Jpn. {\bf 73} (2004) 2257.
\bibitem{otsuki05} J. Otsuki, H. Kusunose, and Y. Kuramoto: J. Phys. Soc. Jpn. {\bf 74} (2005) 2082.
\bibitem{affleck} 
I. Affleck, A.W.W. Ludwig, and B.A. Jones: Phys. Rev. B{\bf 52},  (1995) 9528, and references therein.

\bibitem{kuramoto92} Y. Kuramoto: 
in {\it Transport and Thermal Properties of f-Electron systems}, (G. Oomi, H. Fujii, and T. Fujita eds., Plenum Publ. Corp., 1993) p.237.
\bibitem{shimizu95} Y. Shimizu, O. Sakai, and Y. Kuramoto: Physica B {\bf 206}-{\bf 207} (1995) 135.
\bibitem{koga96} M. Koga and H. Shiba: J. Phys. Soc. Jpn. {\bf 65} (1996) 3007.
\bibitem{hattori05} K. Hattori and K. Miyake: J. Phys. Soc. Jpn. {\bf 74} (2005) 2193.
\bibitem{hoshino09} S. Hoshino, J. Otsuki, and Y. Kuramoto: J. Phys. Soc. Jpn. {\bf 78} (2009) 074719.
\bibitem{hoshino10a} 
S. Hoshino, J. Otsuki, and Y. Kuramoto: 
J. Phys. Soc. Jpn. 79 (2010) 074720.
\bibitem{rubtsov05} A. N. Rubtsov, V. V. Savkin, and A. I. Lichtenstein: Phys. Rev. B {\bf 72} (2005) 035122.
\bibitem{werner06} P. Werner and A. J. Millis: Phys. Rev. B {\bf 74} (2006) 155107.
\bibitem{otsuki07} J. Otsuki, H. Kusunose, P. Werner, and Y. Kuramoto: J. Phys. Soc. Jpn. {\bf 76} (2007) 114707.
\bibitem{georges96} For a review, see: A. Georges, G. Kotliar, W. Krauth, and M. J. Rozenberg: Rev. Mod. Phys. {\bf 68} (1996) 13.
\bibitem{peters07} R. Peters and T. Pruschke: Phys. Rev. B {\bf 76} (2007) 245101.
\bibitem{hoshino10} S. Hoshino, J. Otsuki, and Y. Kuramoto: Phys. Rev. B {\bf 81} (2010) 113108.

\bibitem{watanabe99} S. Watanabe, Y. Kuramoto, T. Nishino, and N. Shibata: J. Phys. Soc. Jpn. {\bf 68} (1999) 159.
\bibitem{tsunetsugu97} For a review, see:
H. Tsunetsugu, M. Sigrist and K. Ueda: Rev. Mod. Phys. {\bf 69} (1997) 809.
\bibitem{otsuki09} J. Otsuki, H. Kusunose, and Y. Kuramoto: J. Phys. Soc. Jpn. {\bf 78} (2009) 034719.

\bibitem{otsuki09a} J. Otsuki, H. Kusunose and Y. Kuramoto: 
Phys. Rev. Letters {\bf 102} (2009)  017202.


\bibitem{iwasa02} 
K. Iwasa {\it et al}: Physica B{\bf 312-313}  (2002) 834.
\bibitem{iwasa03} K. Iwasa {\it et al}: Acta Phys. Pol. B {\bf 34} (2003) 1117.
\bibitem{hao05} 
L. Hao {\it et al}:
Physica B {\bf 359-361} (2005) 871.

\bibitem{park08} J. -G. Park {\it et al}: Phys. Rev. B {\bf 77} (2008) 085102.
\bibitem{iwasa08} K. Iwasa {\it et al}: J. Phys. Soc. Jpn. {\bf 77} (2008) 063706.
\bibitem{schmidt} 
A.R. Schmidt {\it et al}: 
Nature {\bf 465} (2010) 570.


\end{thebibliography}
\end{document}